\documentclass[reprint,
superscriptaddress,
amsmath,amssymb,aps,showkeys,showpacs,
twoside,final,secnumarabic,%raggedbottom,
nofootinbib]{revtex4-2}

\usepackage[paperwidth=205mm,paperheight=290mm,top=17mm,bottom=25mm,
inner=17mm,outer=17mm,
twoside]{geometry}

\usepackage{cmap} % Улучшенный поиск %русских слов в~полученном pdf-файле
\usepackage[T1,T2A]{fontenc}
\usepackage[utf8]{inputenc}
\usepackage[russian,english]{babel}
\usepackage{color}
\usepackage{graphicx}% Include figure files
\usepackage{dcolumn}% Align table columns on decimal point
\usepackage{bm} % bold math
\usepackage[unicode=true,colorlinks=true,linkcolor=magenta, urlcolor=blue, citecolor = blue,breaklinks]{hyperref}
\usepackage{multirow}
\usepackage{url}
\usepackage{breakurl}
\DeclareGraphicsExtensions{.eps}

\newcount\issue
\newcount\Vol
\newcount\numb
\usepackage{fancyhdr} %this packages %provides fancy up and bottom of page
\def\Vol{\textbf{XX}}
\def\numb{x}
\renewcommand{\vec}[1]{\boldsymbol{#1}}
\usepackage{natbib}
\begin{document}

%====== Начало шапки статьи  ============
%\title{JOURNAL SECTION OR CONFERENCE SECTION\\[20pt]
\title{Neutrino Physics\\[20pt]
Neutrino Propagation in Quantum Field Theory at Short and Long Baselines}

\def\addressa{Joint Institute for Nuclear Research, Laboratory of Theoretical Physics}

\author{\firstname{V.A.}~\surname{Naumov}}
\email[E-mail: ]{vnaumov@theor.jinr.ru }
\affiliation{\addressa}
\author{\firstname{D.S.}~\surname{Shkirmanov}}
\email[E-mail: ]{dmitry@shkirmanov.com }
\affiliation{\addressa}

\received{xx.xx.2025}
\revised{xx.xx.2025}
\accepted{xx.xx.2025}

\begin{abstract}
In a quantum field approach to neutrino oscillations, the neutrino is treated as a propagator,
while the external initial and final particle states are described by covariant wave packets.
For the asymptotic behavior on short and long macroscopic baselines, the wave packet modified
neutrino propagator is expressed through asymptotic series in powers of dimensionless Lorentz
and rotation invariant variables.
In both regimes, leading-order corrections violate the classical inverse-square law and lead
to a decrease in the neutrino-induced event rate. The possibility that the so-called reactor
antineutrino anomaly can, at least partially, be explained within this approach is discussed.
\end{abstract}

\par
\keywords{Neutrino, Quantum Field Theory, Wave packet, Inverse square law violation \\[5pt]}
%DOI:

\maketitle
\thispagestyle{fancy}

%====== Начало  статьи  ============

\section{Introduction}\label{intro}

Numerous inconsistencies in the standard quantum-mechanical theory of neutrino oscillations
led to the development of a more consistent approach based on the S-matrix formalism
of quantum field theory (QFT) \cite{Kobzarev:1981ra,*Giunti:1997sk,*Giunti:1991ca,*Grimus:1996av}.
Over the past decades, a number of studies have been devoted to elaborating % refining and expanding
this approach; a list of these studies can be found, e.g., in Ref.\,\cite{Naumov:2020yyv}.
The present study is based on the formalism developed in Refs.\,\cite{Naumov:2009zza,Naumov:2010um}.
Let's briefly recall its key ingredients.

Within the standard ``plane wave'' (PW) $S$-matrix formalism,
single-particle states are defined as Fock states,
$|\vec k, s\rangle =\sqrt{2E_{\vec k}}a_{\vec ks}^\dag|0\rangle$
($E_{\vec k}^2=|\vec k|^2+m^2$), 
which contain no information about the particle's space-time location and cannot be used to
describe the oscillation phenomenon~\cite{Naumov:2009zza,Naumov:2010um,Naumov:2020yyv}.
To account for coordinate dependence, we can use wave-packet (WP) states, which are simply
covariant linear superpositions of Fock states:
\begin{equation}
\label{WavePacketState}
|\boldsymbol{p},s,x\rangle = \int\frac{d\boldsymbol{k}\,\phi(\boldsymbol{k},\boldsymbol{p})e^{i(k-p)x}}
{(2\pi)^32E_{\boldsymbol{k}}}|\boldsymbol{k},s\rangle,
\end{equation}
where $\phi(\vec k, \vec p)$ is a model-dependent scalar function that determines the WP shape
and (``correspondence principle'') transforms into the normalized 3D Dirac $\delta$ function in the PW limit:
$
\phi(\vec k, \vec p){\longmapsto } (2\pi)^32E_{\boldsymbol{p}}\delta(\vec k-\vec p).
$
The amplitude is constructed as the standard QFT amplitude,
$
\langle\mathrm{\bf out}|\mathbb{S}|\mathrm{\bf in}\rangle
\left(\langle\mathrm{\bf in}|\mathrm{\bf in}\rangle\langle\mathrm{\bf out}|\mathrm{\bf out}\rangle\right)^{-1/2}
$,
where {\bf in} and {\bf out} states are, in general, multi-WP states. % not Fock states.
The amplitude depends on the space-time coordinates of all WPs ($\in |\mathrm{\bf in}\rangle$ and $|\mathrm{\bf out}\rangle$),
which interact in the macroscopically separated vertices of a Feynman diagram -- ``source'' and ``detector''
(see Fig.\,\ref{fig_Macrograph_Class_A} in Sect.\,\ref{Sect: The process under study}).
Due to the localised WP states, the amplitude is suitable to describe the processes that
depend on the macroscopic spatial and temporal intervals between the vertices.

\section{Relativistic Gaussian packet}

For our purposes, we use the so-called relativistic Gaussian packet (RGP) model \cite{Naumov:2010um},
in which the ``form factor'' function $\phi(\boldsymbol{k},\boldsymbol{p})$ in Eq.\,\eqref{WavePacketState}
has the form
\begin{align*}
N_{\sigma}
%        \exp\left[\frac{(k-p)^2}{4\sigma^2}\right]
%        \exp\left[\frac{(k_0-p_0)^2-(\vec k -\vec p)^2}{4\sigma^2}\right]}.
         \exp\left[\frac{(E_{\boldsymbol{k}}-E_{\boldsymbol{p}})^2-(\vec k -\vec p)^2}{4\sigma^2}\right]
  \equiv \phi_G(\boldsymbol{k},\boldsymbol{p}).
\end{align*} 
Here $\sigma$ is the momentum spread (dispersion) of RGP and the normalization constant $N_{\sigma}$
is obtained from the linear normalization condition
\begin{equation*}
\int\frac{d\boldsymbol{k}\,\phi_G(\boldsymbol{k},\boldsymbol{p})}{(2\pi)^32E_{\boldsymbol{k}}}=1;
\end{equation*}
% which is consistent with the correspondence principle. 
This is a technical point, but it is useful and fully consistent with the correspondence principle.
The coordinate representation of RGP,
\begin{align*}
\psi_G(\boldsymbol{p},x)
= \int\frac{d\boldsymbol{k}}{(2\pi)^32E_{\boldsymbol{k}}}\phi_G(\boldsymbol{k},\boldsymbol{p})e^{ikx},
% \quad \psi_G(\boldsymbol{p},x)=\frac{K_1({\zeta}m^2/2\sigma^2)}{{\zeta}K_1(m^2/2\sigma^2)}
\end{align*}
can be approximately written as
\begin{align*}
\psi_G(\boldsymbol{p},x)
= \exp\left\{i(px)-\frac{\sigma^2}{m^2}\left[(px)^2-m^2x^2\right]\right\}
\end{align*} 
(so-called ``contracted'' RGP or CRGP), which is valid under two invariant conditions:
$(px)^2 \ll m^4/\sigma^4$ and $(px)^2-m^2x^2 \ll m^4/\sigma^4$.
% These conditions imply that CRGP is narrow in the momentum space.
It's easy to see that $|\psi_G|$ does not depend on time in the proper reference frame.
% i.e., WP does not spread out in the CRGP regime.
% We will use just this approximation in our calculations.
In this study, we will use exactly this approximation.

\section{The process under study}
\label{Sect: The process under study}

\begin{figure}[htb]
\centering\includegraphics[width=0.7\linewidth]{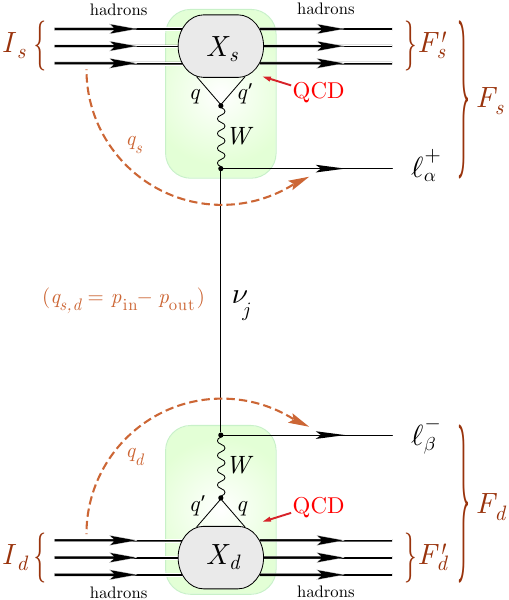}
\caption{Macroscopic Feynman diagram representing the process under discussion.
         Initial and final states in source and detector vertices are wave packets.
         Notations are explained in the text. For details, see Ref.\,\cite{Naumov:2020yyv}.
        }
\label{fig_Macrograph_Class_A}
\end{figure}

Figure \ref{fig_Macrograph_Class_A} shows a generic macroscopic Feynman diagram
describing the class of processes under consideration.
Here $X_s$ and $X_d$ denote macroscopically separated regions in space and time
in which intermediate neutrinos are produced and detected, respectively.
% These regions are \textbf{macroscopically} separated in space-time.
$I_{s,d}$ denote initial particles (WPs) at the source and detector vertices, respectively;
$F_{s,d}$ and $F'_{s,d}$ denote the sets of final WPs (the notation is clear from the figure). 
The ``oscillations'' (= flavor transitions) arise from the interference of the Feynman diagrams
with different mass (virtual) eigenfields $\nu_j$ ($j=1,2,3$) in the intermediate states.
In other words, in this formalism, neutrino oscillations are the result of interference, not mixing.
The amplitude of the process is determined by the Lagrangian of the Standard Model,
extended to include neutrino masses:
\begin{eqnarray*}
\mathcal{L}_W(x) = -\frac{g}{2\sqrt{2}}\left[j_\ell(x)W(x)+j_q(x)W(x)+\mathrm{H.c.}\right].
\end{eqnarray*}
Here $g$ is $SU(2)$ coupling constant; $j_\ell(x)$ and $j_q(x)$ are the lepton and quark weak charged currents:
\begin{eqnarray*}
j_\ell^{\mu}(x) = \sum_{{\alpha}i}V_{{\alpha}i}^*\,\overline{\nu}_i(x)\gamma^\mu(1-\gamma_5)\ell_{\alpha}(x), \\
j_q^{\mu}(x) = \sum_{qq'}V_{qq'}^{'*}\,\overline{q}(x)\gamma^\mu(1-\gamma_5)q'(x).
\end{eqnarray*} 
Here $\alpha=e,\mu,\tau$, $i=1,2,3$, $q=u,c,t$, $q'=d,s,b$, $V_{{\alpha}i}$ and $V_{qq'}$ are
elements of the neutrino and quark mixing matrices (PMNS and CKM), respectively;
$\ell_{\alpha}(x)$ are the lepton field, $q(x)$ and $q'(x)$ are the quark fields. 
% and $O^\mu=\gamma^\mu(1-\gamma_5)$.
For our aims, we retain only the leading non-vanishing term of the perturbation theory.

\section{Inverse square law violation}
\label{Sect:ISLV}

It can be shown that space-time dependence of the amplitude is defined by the 
neutrino propagator modified by the external WPs:
% external wave-packet-modified $\nu$ propagator:
\begin{equation}\label{the propagator}
J(X)=\int\frac{d^4{q}}{(2\pi)^4}\;\frac{\widetilde\delta_s(q-q_s)\widetilde\delta_d(q+q_d)(\hat{q}+m)e^{-i{qX}}}{{q}^2-m^2+i\epsilon}.
\end{equation}
Here ${\widetilde\delta_{s,d}}$ are the ``smeared'' $\delta$-functions responsible
for approximate 4-momentum conservation at the source and detector vertices,
$X=(T,\vec{L})$ is the space-time separation between the source and detector vertices;
$q_{s,d}$ are the 4-momentum transfers at the source/detector; and $m$ ($\equiv m_j$) is the neutrino mass.

\begin{figure}[htb]
\centering\includegraphics[width=0.99\linewidth]{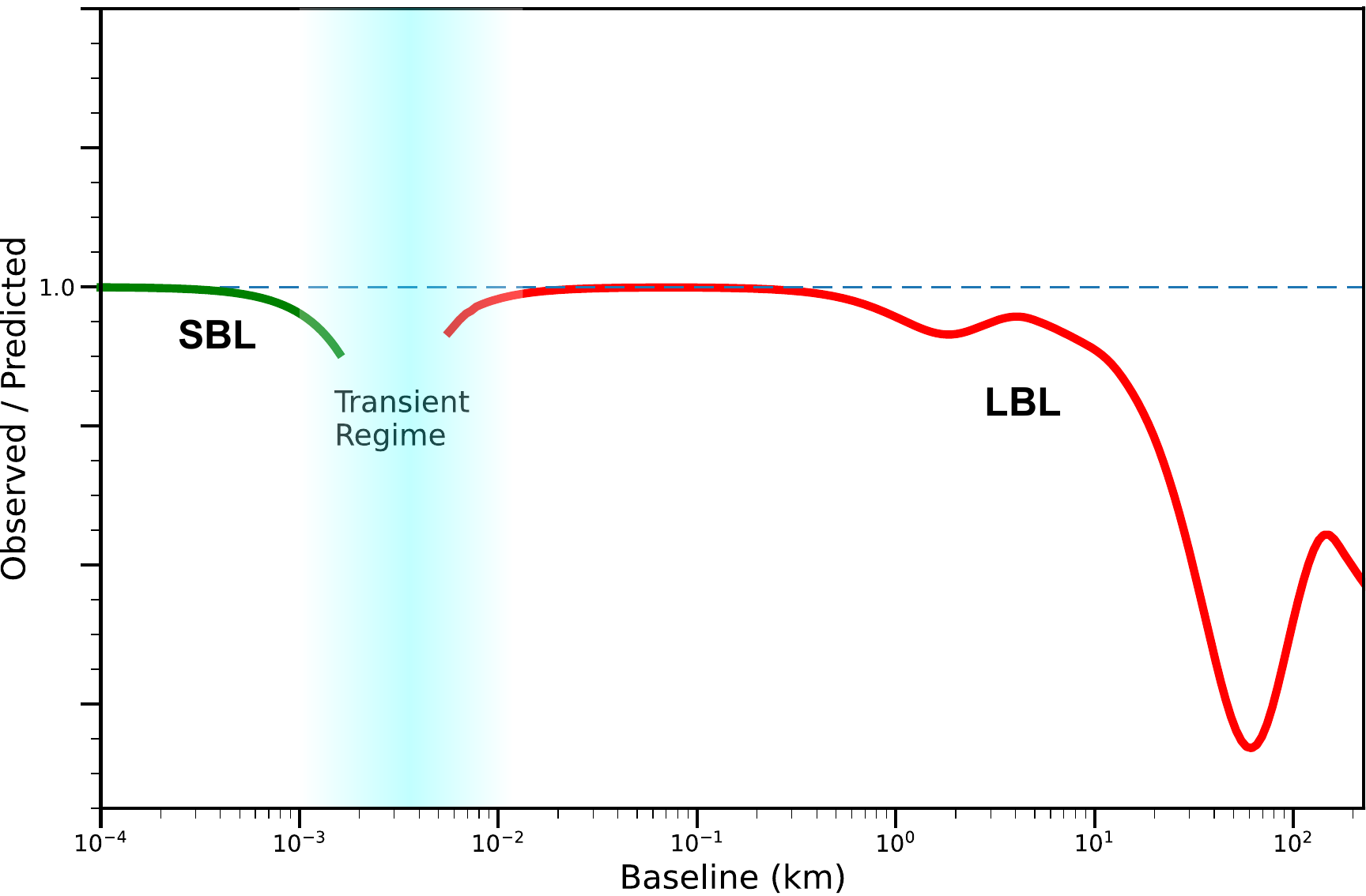}
\caption{Graphical representation of the ISL effect 
         (observed-to-predicted ratio vs.\ distance).}
\label{fig:ISLV}
\end{figure}

Integral \eqref{the propagator} was studied in detail in two opposite asymptotic regimes:
short baselines~\cite{Naumov:2022kwz}
(${{1}/{\Sigma_{\text{SBL}}^2} \ll L^2 \ll {{|\vec q|^2}}/{\Sigma_{\text{SBL}}^4}} $)
and long baselines ~\cite{Naumov:2013bea}
($ L^2 \gg {{|\vec q|^2}}/{\Sigma_{\text{LBL}}^4}$), 
where $\Sigma_{\text{SBL}}$ and $\Sigma_{\text{LBL}}$ are the effective momentum scales
for the short and long baseline asymptotics, respectively.
These scales are defined by the momentum spreads $\sigma_\varkappa$, masses $m_\varkappa$,
and momenta $\vec{p}_\varkappa$ of all {\bf in} and {\bf out} WPs
$\varkappa \in I_s \oplus I_d  \oplus F_s  \oplus F_d$ % (in source and detector)
in the diagram in Fig.~\ref{fig_Macrograph_Class_A}.
% Finally, $\vec q$ is the neutrino momentum.
Since the solutions are quite cumbersome, we will provide only the briefest summaries here.
For the SBL asymptotics, the solution is a triple asymptotic series in powers
of small invariant parameters dependent of the deeply virtual neutrino 4-momentum and 4-vector $X$.
For the LBL asymptotics, the 3D part (over $\vec{q}$) of the 4D integral \eqref{the propagator} can
be evaluated using the extended Grimus-Stockinger theorem~\cite{Naumov:2013bea,Korenblit:2014uka}.
% (the solution is an asymptotic series ). 
The remaining integral over $q_0$ for the LBL asymptotics can be evaluated using the saddle-point method.
Using these solutions, one can show that in the \textbf{ultrarelativistic}
approximation, the number of neutrino events detected during the detector exposure time  $\tau_d$ 
can be (somewhat symbolically) written as
\begin{multline}
\label{AveragedProbability_Simplified_mod4}
\tau_d\sum\limits_{\text{spins}}\int{d}\vec{x}\int{d}\vec{y}\int{d}\mathfrak{P}_s\int{d}\mathfrak{P}_d
\int{d}\vert\vec{q}\vert \\  \times
\frac{\mathcal{P}_{\alpha\beta}\left(\vert\vec{q}\vert,\vert\vec{y}-\vec{x}\vert\right)}{4(2\pi)^3{\vert\vec{y}-\vec{x}\vert^2}}
\!\times\!\left(1-{\text{ISLV corrections}}\right).
\end{multline}
Here, the differential forms $d\mathfrak{P}_{s,d}$ are defined as
\[
\begin{aligned}
d\mathfrak{P}_s = 
&\ \prod\strut_{a{\in}I_s}~\frac{ d\vec{p}_a f_a(\vec{p}_a,s_a, \vec x)}{(2\pi)^32E_a} \\ \times
&\ \prod\strut_{b{\in}F_s}~\frac{ d\vec{p}_b }{(2\pi)^32E_b}  (2\pi)^4{\delta}_s(q-q_s)\vert{M}_s\vert^2, \\
d\mathfrak{P}_d = 
&\ \prod\strut_{a{\in}I_d}~\frac{ d\vec{p}_a f_a(\vec{p}_a,s_a,\vec y)}{(2\pi)^32E_a}  \\ \times
&\ \prod\strut_{b{\in}F_d}~\frac{[d\vec{p}_b]}{(2\pi)^32E_b}  (2\pi)^4{\delta}_d(q+q_d)\vert{M}_d\vert^2;
\end{aligned}
\]
% Here, $\tau_d$ is detector exposure time; 
integrations over $\vec{x}$ and $\vec{y}$ are performed over the spatial volumes of the source and detector
physical devices, respectively;
$f_a(\vec{p}_a,s_a,\vec{x})$ is the distribution function of particles of type $a$ in the source
(for $d\mathfrak{P}_s$) and detector (for $d\mathfrak{P}_d$);
$E_{a,b}$ are the energies of the corresponding particles;
$\delta_{s,d}$ are ``smeared'' delta functions (not identical to $\widetilde\delta_{s,d}$) responsible
for an approximate energy-momentum conservation in the source and detector, respectively;
$M_{s,d}$ are the standard QFT matrix elements describing interaction in the source and detector respectively.
Square brackets in $d\mathfrak{P}_d$ indicate that there is no integration over the final particles in the detector.
The function $\mathcal{P}_{\alpha\beta}\left(\vert\vec{q}\vert,\vert\vec{y}-\vec{x}\vert\right)$ is a generalization of the QFT
oscillation probability, which reproduces the standard quantum mechanical formula with some modifications that are not essential in the context of this study.
The term ${\vert\vec{y}-\vec{x}\vert^2}$ in the denominator of Eq.\,\eqref{AveragedProbability_Simplified_mod4}
accounts for the standard inverse-square law (ISL), which states that the neutrino event rate is proptional to
$1/L^2$, where $L=|\vec{L}|$ is the distance between the source and detector devices.
Finally, the ``ISLV corrections'' in Eq.\,\eqref{AveragedProbability_Simplified_mod4} denotes the asymptitic
expansion that depends on $\vert\vec{x}-\vec{y}\vert$ and is responsible for the ISL violation (ISLV).
In the leading order, these corrections yield  the following distance dependence of the event rate for the
long and short baselines, respectively:
\begin{equation*}
\begin{aligned} 
\frac{dN_\nu}{dt} \propto &\ {\frac{1}{L^2}}
\left[1-\frac{L_{0}^2}{{L^2}}+\ldots\right], \quad  L^2 \gg \frac{{E_\nu^2}}{\Sigma_{\text{LBL}}^4}, \\ 
\frac{dN_\nu}{dt}  \propto &\ {\frac{1}{L^2}}
\left[1-\frac{{L^2}}{L_0'^2}+\ldots\right], \quad \frac{1}{\Sigma_{\text{SBL} }^2}
                                          \ll L^2 \ll  \frac{{E_\nu^2}}{\Sigma_{\text{SBL}}^4}.
\end{aligned}
\end{equation*}
Clearly, the factor $1/L^2$ (for both asymptotics) represents the classical ISL and the 
power corrections in the square brackets violate the ISL. 
As can be seen, for both regimes, the ISLV corrections decrease the neutrino event rate
in the detector. The parameters $L_0$ and $L_0'$ define the scales at which the ISLV occurs.
The order of magnitude of these parameters can be very roughly estimated as:
\begin{equation}
\label{L0}
L_0\sim L_0'   \sim    20 \left\langle \left(\frac{E_\nu}{1\;\text{MeV}}\right)
\left[\frac{\sigma_{\text{eff}}(E_\nu)}{1\;\text{eV}}\right]^{-2}\right\rangle\;\text{cm}.
\end{equation}  
Here, $\sigma_{\text{eff}}\sim \Sigma_{\text{SBL}} \sim \Sigma_{\text{LBL}}$ and
$E_{\nu}$ is the virtual neutrino energy; 
the formalism does not formally exclude the possibility that $\Sigma_{\text{SBL}}\gtrsim\Sigma_{\text{LBL}}$.
% One can see that the ISLV could be observable at distances measured in meters for reasonable values
% of $\sigma_{\text{eff}}$. 
It is clear that the ISLV for reactor antineutrinos can be observed at distances of the order of meters,
for reasonable values of the parameter $\sigma_{\text{eff}}(E_\nu)$.
In Ref.~\cite{Naumov:2021vds} and in our earlier papers \cite{Naumov:2015hba,*Naumov:2017pgt},
the possibility is discussed that the ISLV could be responsible (at least partly) for the
long-standing reactor antineutrino anomaly and maybe for the anomalies in neutrino experiments with
the artificial neutrino sources at Ga-Ge solar neutrino detectors (see Ref.\,\cite{Naumov:2021vds}
and references therein).
When the ISLV effects are taken into account, the principal behavior of the neutrino event rate
in the detector as a function of distance may appear as symbolically shown in Fig.~\ref{fig:ISLV}.
For both LBL and SBL asymptotics, the theory predicts a neutrino deficit at short distances.
% The transient regime in the middle cannot be described by obtained asymptotic series.
The transient regime cannot be described by the two obtained asymptotic expansions and requires
a different mathematical analysis.

\section{Reactor Experiments and Inverse Square Law Violation}

\begin{figure}[htb]
\centering 
\includegraphics[width=\linewidth]{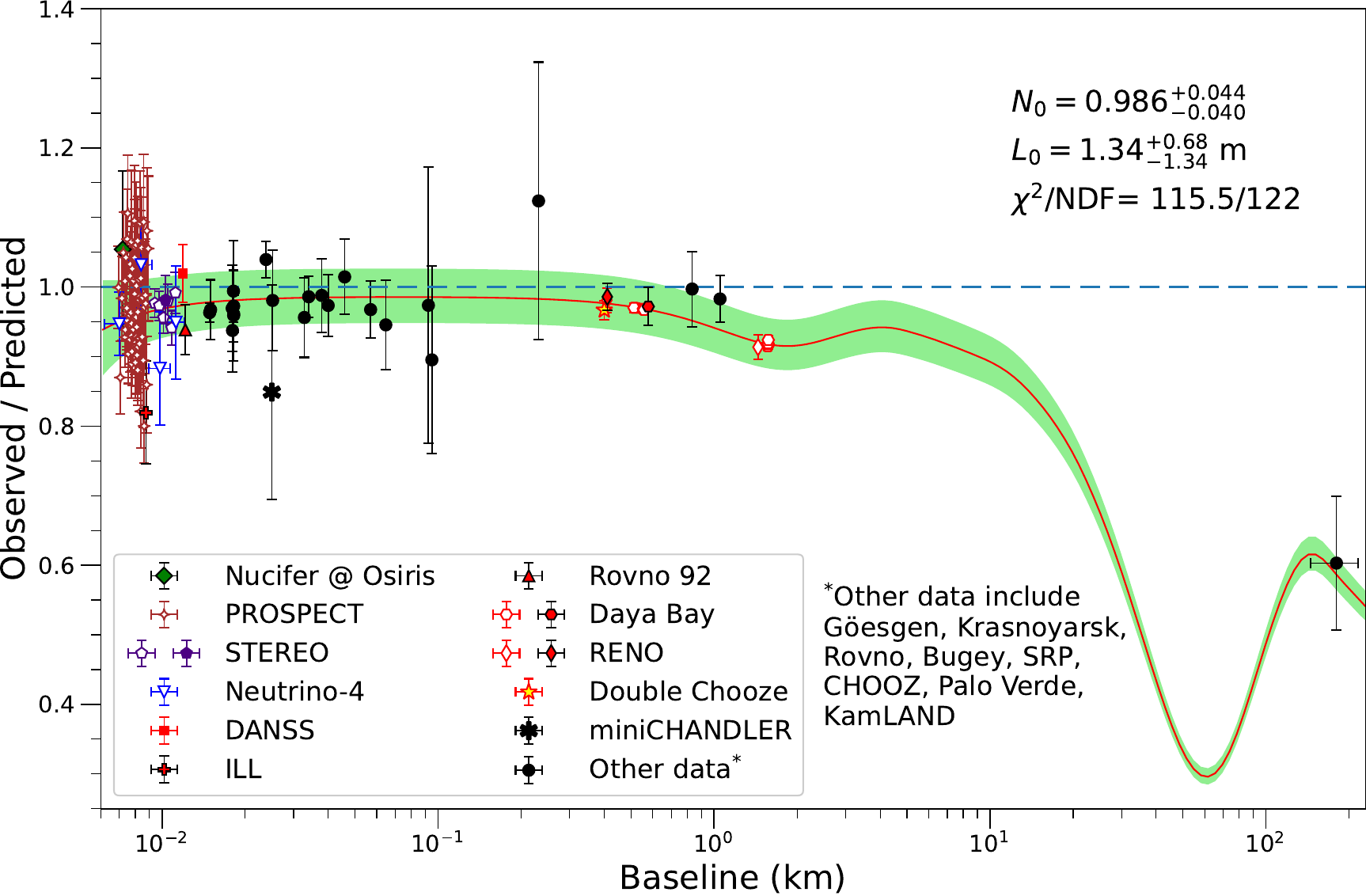}   
\caption{Ratio of observed to predicted rates for the KI $\overline\nu_e$ spectrum~\cite{Kopeikin:2021ugh}.
         Solid line represents the best-fitting theoretical model~\eqref{eq:fit_prediction}, 
         and the band corresponds to the 68\% CL. 
         Filled points represent  absolute measurements, and open ones are for the relative measurements
         normalized to the best-fit curve; references to experiments are given in Ref.\,\cite{Naumov:2021vds}. 
         The experimental errors shown do not include the overall normalization uncertainty, which is at least 2.7\%.
        }
\label{fig:fit_results_KI}
\end{figure}

To search for the inverse-square violation effect, we analyzed reactor
$\overline\nu_e$ experiments. As part of our analysis, we estimated systematic
error correlations in the reactor data set. The theoretical model used for the
analysis has the following form:
\begin{equation}
\label{eq:fit_prediction}
T(L;N_0,L_0) = 
% N_0 T(L;L_0)= 
N_0 \cdot \langle P^{3\nu}_\text{surv}(L) \rangle \cdot {\left(1-\frac{L_0^2}{L^2}\right)}.
\end{equation}
Here, $L_0$ is a free parameter responsible for the inverse-square violation effect,
$N_0$ is a free normalization parameter introduced to account for $\overline\nu_e$ flux uncertainty; and
\begin{equation*}
\langle P^{3\nu}_\text{surv}(L)\rangle = 
\dfrac{\displaystyle\int dE \sum\limits_k f_k P^{3\nu}_\text{surv}(L,E)\sigma(E) S_k(E)}
      {\displaystyle\int dE \sum\limits_k f_k \sigma(E) S_k(E)}.
\end{equation*}
Here, $f_k$ is the fraction of the main fissile isotope contributing to the $\overline{\nu}_e$ flux with an energy spectrum $S_k(E)$, %\cite{Mueller:2011nm}
$\sigma(E)$ is the IBD cross section~\cite{Ivanov:2013cga}, and
$P^{3\nu}_\text{surv}(L,E)$ is the $\overline{\nu}_e$ survival probability in the $3\nu$ mixing scheme:
\begin{multline*}
P^{3\nu}_\text{surv}(L,E) =  1-\cos^4\theta_{13}\sin^2\left(2\theta_{12}\right)\sin^2\Delta_{21} \\
-\sin^2\left(2\theta_{13}\right)\left(\cos^2\theta_{12}\sin^2\Delta_{31} + \sin^2\theta_{12}\sin^2\Delta_{32}\right),
\end{multline*}
with $\Delta_{ij}= 1.267\Delta m_{ij}^2 L/E$. We performed our analysis for
different models of $\overline\nu_e$ spectra to verify the consistency of the results.
Figure~\ref{fig:fit_results_KI} shows the fit results for the Kurchatov Institute (KI) $\overline\nu_e$
spectrum~\cite{Kopeikin:2021ugh}.
The best-fit values of the parameter are shown in the figure.
A non-zero best-fit value of the parameter $L_0$ indicates that the reactor data indeed exhibits a slight ISLV.
At the same time, the accuracy of the reactor data does not allow for a reliable determination of the lower bound
for the parameter $L_0$. 
%Thus, for now, we can only speak of some indication of the effect.
Different models (Huber-Mueller~\cite{Huber:2011wv,Mueller:2011nm}, Fallot et al.~\cite{Fallot:2012jv})
of $\overline\nu_e$ spectra yield similar results: while the best-fit value of $N_0$ depends significantly
on the spectrum model, the best-fit value of $L_0$ changes only slightly, well within the uncertainties
of the analysis.
For example, for the Huber-Mueller $\overline\nu_e$ spectrum, we obtained the following
best-fit parameter values: $N_0 = 0.948^{+0.042}_{-0.039}$, $L_0 = 1.34^{+0.68}_{-1.34}$, $\chi^2/\text{NDF}= 115.5/122$. As can be seen, the
$N_0$ parameter is inconsistent with unity, indicating a discrepancy
between the experimental data and the Huber-Mueller spectrum. For the
Fallot spectrum, the best-fit parameter values are $N_0 = 1.005^{+0.044}_{-0.041}$, $L_0 = 1.21^{+0.73}_{-1.21}$,
$\chi^2/\text{NDF} = 116.9/122$. Clearly, the Fallot spectrum does not
require flux normalization.

% \vspace*{-5.0mm} % Does not work!

\section{Conclusions and Discussion}

The QFT approach to neutrino oscillations predicts that the classical
inverse-square law could be violated. The scale at which this violation occurs
depends on the momentum spreads of the {\bf in} and {\bf out} wave packets, which
are parameters of the theory; therefore, the actual distance at which the ISLV occurs
cannot be derived from ``first principles''.
The available reactor data hint that ISLV has already been observed, but the
best-fit value of $L_0$ is formally compatible with zero. New experiments with
intense $\nu/\overline{\nu}$ sources and sectioned or movable detector(s)
(such as BEST-2, SOX/CeSOX, CeLAND) are required in order to test the theory.

It is possible that ISLV is also responsible for anomalies in experiments with
artificial (anti)neutrino sources as well, although the numerical value of $L_0$ would
differ for these experiments, since $L_0$ depends on neutrino energy according to Eq.\,\eqref{L0}. 
We would also like to add that if new data (e.g., from DANSS) rule out ISLV,
it would not mean a confutation of the QFT approach. Rather, it would only
indicate that $\sigma_{\text{eff}}$ is above the experimental sensitivity threshold.

% \vspace*{-5.0mm} % Does not work!

\section{CONFLICT OF INTEREST}
The authors declare that they have no conflicts of interest.

% most likely will not want to use it. \nocite{*}

%%%%%%%%%%%%%%%%%%%%%%%%%%%%%%%%
% USE thebibliography
%%%%%%%%%%%%%%%%%%%%%%%%%%%%%%%%

\bibliography{references}
%\begin{thebibliography}{}
%% book
%\bibitem{bib1}
%A. van der Woude, Int. Review of Nuclear Physics, \textit{Electric and Magnetic Giant Resonances in Nuclei} (World Scientific, Singapore, 1991), pp. 99–232.
%
%% book in Russian
%\bibitem{bib2}
%O. V. Bogdankevich and F. A. Nikolaev, \textit{Bremsstrahlung Beam (Features of the Technique of Physical Research on Electron Accelerators)} (Atomizdat, Moscow, 1964). (In Russian).
%
%% dissertation
%\bibitem{bib3}
%L. Z. Dzhilavyan, "Photonuclear research in the field
%of giant resonances in forward and backward reaction", Doctoral Dissertation in Physics and Mathematics (Inst. for Nuclear Research, Russ. Acad. Sci., 2017).
%
%%article
%\bibitem{bib4}
%E. De Sanctis, M. Anghinolfi, G. P. Capitani, et al.,
%Phys. Rev. C 34, 413 (1986). https://doi.org/10.1103/PhysRevC.34.413.
%
%\end{thebibliography}

\end{document}